# Understanding the atomically precise evolution of the miscibility of newly prepared face-centered cubic W–Cu nanoalloys and its asymmetry

Yongxin Zhang[1,2,3†], Weihan Zhang[4†], Luneng Zhao[5], Zixiang Zhao[1,2,3], Siqi Lu[1,2,3], Yangrui Liu[6], Dongsheng Song[6], Changzheng Wei[1,2,3], Zhentao Pang[7], Yifeng Ren[1], Junfeng Gao[5], Weiwei Gao[2,3], Di Wu[1,2,3], Jijun Zhao[8], Kuojuei Hu[1,2,3*], Wei Ji[4*], Yu Deng[7*], Binghui Ge[6*], and Fengqi Song[1,2,3*]

[1]National Laboratory of Solid State Microstructures, Collaborative Innovation Center of Advanced Microstructures, and School of Physics, Nanjing University, Nanjing 210093, China.

[2]Institute of Atom Manufacturing, Nanjing University, Suzhou 215163, China.

[3]Nanjing Institute of Atomic Scale Manufacturing, Nanjing 211800, China.

[4]Department of Physics and Beijing Key Laboratory of Optoelectronic Functional Materials and Micro–Nano Devices, Renmin University of China, Beijing 100872, P. R. China.

[5]Key Laboratory of Materials Modification by Laser, Ion and Electron Beams, School of Physics, Dalian University of Technology, Dalian 116024, China.

[6]State Key Laboratory of Opto-Electronic Information Acquisition and Protection Technology, Leibniz International Joint Research Center of Materials Sciences of Anhui Province, Institutes of Physical Science and Information Technology, Anhui University, Hefei, 230601, China.

[7]National Laboratory of Solid State Microstructures, Jiangsu Key Laboratory of Artificial Functional Materials, College of Engineering and Applied Sciences, Collaborative Innovation Center of Advanced Microstructures, Nanjing 210023, China.

[8]Guangdong Provincial Key Laboratory of Quantum Engineering and Quantum Materials, School of Physics, South China Normal University, Guangzhou 510006, China.

†These authors contributed equally to this work. *Correspondence and requests for materials should be addressed to F. S. (email: songfengqi@nju.edu.cn), W. J. (email: wji@ruc.edu.cn), K. H. (email: kuojueihu@nju.edu.cn), B. G. (email: bhge@ahu.edu.cn), and Y. D. (email: dengyu@nju.edu.cn).

**Abstract**

According to classical Miedema theory, reducing crystals to the order of nanometer sizes might greatly modulate the mixing enthalpy of elements, thus enabling the invention of a lot of new bulk-immiscible alloys. Although numerous alloys with higher mixing enthalpies remain unexplored, this strategy is approaching its limit, as reflected by the critical diameter of recent alloys of 1.8 nm, which corresponds to ~150 atoms and hardly provides a crystalline order. Future development requires not only even smaller atomic-scale control but also a new surface energy-saving mechanism. Here, we report the formation of W–Cu nanoalloys with a very large miscibility gap in the bulk via the use of an atomically size-selected cluster beam source as an example. The face-centered cubic (FCC) structure was demonstrated through electron diffraction, which indicated a lattice constant of 3.88 Å for $W_{0.85}Cu_{0.15}$ nanoalloys (~2280 atoms). In this comprehensive study that covers a large parameter space of W/Cu compositions and numbers of atoms, an asymmetric miscibility nanophase diagram in which W-rich compositions favor mixing and the critical size is approximately 6000 atoms, which far exceeds the approximately tens of atoms predicted via classical theory, was obtained for the first time. Density functional theory (DFT) calculations revealed a mutual strain-induced mechanism that simultaneously lowers the surface energies while reducing the size to the atomic scale. This approach paves the way for the development of new high-performance nonequilibrium phase alloys.

**Introduction**

Creating new alloys in thermodynamically immiscible systems is a key area of research within the field of materials science.[1-3] Alloy materials are ubiquitous and play pivotal roles in various fields, including biomedicine, aerospace, and additive manufacturing.[4-8] The enhanced performance of these materials is attributed to their synergistic effects, which are mediated by the chemical, mechanical, and electronic interactions among their constituent elements.[1,9-11] However, owing to differences in atomic radii, valence states, electronegativities, and crystal structures, thermodynamically immiscible systems exhibit miscibility gaps in their phase diagrams, which precludes the formation of alloys through conventional techniques.[12-16]

On the basis of classical Miedema theory, mixing is more challenging in systems with higher mixing enthalpies ($\Delta H_{mix}$). Recent studies have shown that reducing the size to the order of nanometers can decrease $\Delta H_{mix}$, thereby leading to significant progress in the formation of new alloys from immiscible systems.[3,11,17-20] For example, new alloys such as Ag–Cu, Ta–Cu, Au–Rh, Au–Ni, and Ag–Ni, whose $\Delta H_{mix}$ values are 2 kJ/mol, 2 kJ/mol, 7 kJ/mol, 7 kJ/mol, and 15 kJ/mol, respectively, have demonstrated mixing.[1,2,9-12,16,18,21-24] In recent work by Prof. PD Yang, Au–Rh nanoparticles were successfully mixed at the extreme limit size below 1.8 nm (approximately 150 atoms).[10]

However, structures that contain fewer than 147 atoms cease to be crystalline and instead exhibit cluster–shell geometries from a cluster perspective, which indicates that the size limit has been reached in recent works. Consequently, the large unexplored field of new alloys with even higher $\Delta H_{mix}$ values, such as W–Cu, Mo-Ag, and W-Ag, which have $\Delta H_{mix}$ values of 22 kJ/mol, 37 kJ/mol, and 43 kJ/mol, respectively, poses substantial synthesis challenges. Adopting the W–Cu system, which is an ideal material for fusion reactor divertors, as an example of a high-$\Delta H_{mix}$ system with significant potential for various applications, according to Miedema theory, a nanocrystalline structure may appear at the scale of 10 atoms. This finding indicates that future exploration of high-enthalpy alloys requires a new mechanism for reducing the surface energy of the nanocrystals, in addition to more precise and smaller control of the diameter [18,25,26].

Here, we report the successfully preparation of FCC-structured $W_{0.85}Cu_{0.15}$ nanoalloys by using a size-selected cluster beam source.[27-33] A systematic investigation with atomic precision demonstrated an asymmetric nanophase diagram that favors W-rich phases and a critical size of approximately 6000 atoms for the first time. A new mechanism is proposed on the basis of both DFT calculations and quantitative model fitting, where mutual strain simultaneously reduces the FCC (111) surface energies of both W and Cu, thus compensating for the destabilization energy terms and enabling the formation of unexpected alloy nanocrystals.

**Successful preparation of atomically precise $W_{0.85}Cu_{0.15}$ miscible nanoalloys**

To precisely generate W–Cu clusters with predetermined numbers of atoms and compositions, we used a magnetron sputtering gas-phase condensation cluster beam source equipped with a time-of-flight mass selector (**Figure 1a**).[34] First, W and Cu atoms condensed and nucleated during flight to form W–Cu clusters of various sizes. Owing to the difference in pressure, the clusters were subsequently cooled to extremely low temperatures at a rate of approximately $10^5$ K/s after passing through a nozzle.

Thereafter, mass-selected W–Cu cluster beams were obtained via a time-of-flight mass selector (the mass spectrum is shown in **Supplementary Figure 1**), and they softly landed on a gold grid that coated with a carbon film; they exhibited an energy lower than the binding energy of each atom.[35] All these processes were conducted in a vacuum environment with a pressure of $10^{-5}$ Pa. Finally, the gold grid that was coated with a carbon film and featured deposited W–Cu clusters was transferred into a vacuum transfer tomography holder by using a glove box that was connected to the cluster beam source, which was filled with inert gas, to ensure that the clusters remained unoxidized.

We selected W and Cu as a model system that is completely immiscible at temperatures below 2000°C and exhibits a substantial miscibility gap in the corresponding phase diagram.[13] W had a body-centered cubic (BCC) structure, whereas Cu had an FCC structure (**Figure 1b**). The generated mass-selected clusters were characterized via aberration-corrected high-angle annular dark-field scanning transmission electron microscopy (AC–HAADF–STEM). The composition and distributions of the W and Cu atoms were determined through energy-dispersive X-ray spectroscopy (EDS). The significant difference in the atomic number between W and Cu allowed for their easy distinction in HAADF–STEM images, as the image intensity was proportional to the atomic number ($I=kNZ^{\alpha}$).[36-38] Notably, TEM sample holders typically comprise Cu, which can introduce a large Cu background-induced signal error during EDS measurements. To circumvent this issue, we employed a vacuum transfer tomography holder with a titanium tip, thereby effectively avoiding this problem (**Supplementary Figure 2**). Moreover, the sample holder employed provided excellent oxidation resistance for the clusters (**Supplementary Figure 3**).

A typical EDS spectrum of $W_{0.85}Cu_{0.15}$ clusters is shown in **Figure 1c**, where the peaks of W and Cu can be clearly identified. To better understand the structure of the $W_{0.85}Cu_{0.15}$ clusters, we systematically characterized such clusters with a mass of 378488 atomic mass units (a.m.u., approximately 2280 atoms) through a series of techniques (**Figure 1d–g** and **Supplementary Figure 4**). Low-magnification and high-magnification HAADF–STEM images are shown in **Figure 1d** and **Figure 1f**, respectively; these images show the excellent size uniformity of the $W_{0.85}Cu_{0.15}$ clusters and their distinct lattice structures in a certain orientation. Owing to the small sizes of the clusters (approximately 4 nm in diameter) and their random orientations, in a wide field of view, the selected-area electron diffraction (SAED) pattern of $W_{0.85}Cu_{0.15}$ clusters exhibited the expected polycrystalline diffraction rings, which are slightly larger than the rings common in bulk materials (**Figure 1e**). The SAED pattern was indexed by using the parameters of the FCC structure. The suitable match indicated the predominant presence of the FCC structure. However, the interplanar spacing that was measured from **Fig. 1f** was insufficient for determining the cluster structure. Nanobeam electron diffraction (NBED) was subsequently performed on the same batch of samples, and the results revealed that the $W_{0.85}Cu_{0.15}$ clusters exhibited an FCC structure with a [110] zone axis (**Figure 1g**). The lattice parameter was determined to be approximately 3.88 Å through measurement and calculation. Notably, the FCC structure indicated by the $W_{0.85}Cu_{0.15}$ clusters was not the BCC structure of W common in bulk form; rather, W could only adopt the FCC structure in computational simulations. The lattice constant of $W_{0.85}Cu_{0.15}$ occurred between the lattice constants of W with a pure FCC structure (a = 4.06 Å) and Cu with an FCC structure (a = 3.58 Å) that were extracted from previous studies.[39,40] All these characterizations collectively demonstrated that we generated $W_{0.85}Cu_{0.15}$ single-phase solid solution nanoalloys with a mass of 378488 a.m.u. by using an atomically size-selected cluster beam source, thus achieving atomic-scale miscibility between W and Cu.

**Evolution of the miscibility with respect to the number of atoms and composition**

To investigate miscibility as a function of the composition and total number of atoms, we generated a series of W–Cu clusters with total numbers of atoms that ranged from approximately 300 to 10,000 and W contents that ranged from 20% to 85%. We considered five elements ($W_xCu_{1-x}$, where x=0.85, 0.65, 0.5, 0.3, or 0.2; **Fig. 1c**, **Figure 2a**, and **Supplementary Figure 5**) and obtained clusters with at least five distinct masses for each composition. The EDS mapping and line-scanning results for a miscible $W_{0.85}Cu_{0.15}$ cluster and an immiscible $W_{0.2}Cu_{0.8}$ cluster are shown in **Figures 2b** and **2c**, respectively, both of which had the same total mass of 935272 a.m.u.

To analyze the distribution of W in the clusters, we developed a random walk algorithm for detecting bright areas in STEM images and identifying W-enriched regions. In **Figure 2d**, we selected $W_{0.5}Cu_{0.5}$ clusters with five distinct sizes (with masses of 56856, 103224, 169832, 378488, and 935272 a.m.u., which corresponded to 458, 832, 1370, 3052, and 7542 atoms, respectively) for the detection of W-enriched regions. In the smallest cluster (left), W was dispersed under projection, whereas the other four larger clusters exhibited distinct phase separation conditions. Additionally, as shown in **Figure 2e**, we selected five clusters with a mass of 169832 a.m.u. and distinct compositions ($W_xCu_{1-x}$, where x=0.85, 0.65, 0.5, 0.3, or 0.2) for the detection of W-enriched regions. In the clusters with higher W contents, such as $W_{0.85}Cu_{0.15}$ and $W_{0.65}Cu_{0.35}$, W was dispersed under projection, whereas the three clusters with lower W contents showed distinct phase separation. These analyses collectively demonstrated that with decreasing number of atoms and increasing W content, phase-separated Cu and W atoms were prone to mixing within a cluster.

**DFT modeling of size- and composition-dependent miscibility**

As mentioned previously, in bulk form, W atoms crystallize preferentially in a BCC lattice, whereas Cu atoms favor stacking in an FCC lattice (**Fig. 1b**). This structural disparity, along with a large positive $\Delta H_{mix}$, leads to W and Cu being conventionally immiscible.[41-44] For nanocrystals and nanoparticles, a high surface-to-volume ratio is a characteristic feature that increases the surface energy, which is a significant component of the total energy. The fractions of bulk, surface, edge, and corner atoms in a truncated-cube nanocrystal model for FCC metals are shown in **Fig. 3a**. As shown in **Fig. 3a**, the bulk and surface portions were of the highest and secondary importance in most cases, except for those nanocrystals comprising fewer than 300 atoms, for which the edge and corner portions were more significant. Given this analysis of the atomic portions, we subsequently examined the energies of the bulk and surface atoms with distinct W–Cu compositions.

The average bond length of a metal alloy typically depends on the composition of the alloying elements.[45-47] We thus used the virtual crystal approximation (VCA) method to interpolate average bond lengths across the entire $W_xCu_{1-x}$ composition range of interest (refer to **Supplementary Figure 6** for more details).[46] The equilibrium bond length of W (2.723 Å or 2.819 Å) was greater than that of Cu (2.457 Å or 2.524 Å) in the BCC and FCC lattices, respectively. The alloy always had an average bond length between those of pure Cu and W with monotonic composition dependence. Specifically, the alloy experienced mutual strain upon alloying such that the W–W (Cu–Cu) bond length was compressed (stretched). To quantitatively assess the reliability of the VCA-predicted bond lengths, we compared our calculated values with our experimental measurements. For miscible $W_{0.85}Cu_{0.15}$ alloy nanocrystals, the VCA model yielded an average lattice constant of 3.89 Å, which is highly consistent with the experimentally measured value of 3.88 Å.

On the basis of the interpolated bond lengths, the energies of the bulk W (gray symbols) and Cu (orange symbols) crystals encompassing BCC (solid dots) and FCC (grid squares) lattices are shown in **Fig. 3b.** Upon W–Cu alloying, both W atoms with compressed bond lengths and Cu atoms with stretched bond lengths exhibited significantly increased energies in both the BCC and FCC lattices. Moreover, the W crystals more rapidly destabilized for alloys with W contents less than 30%. The observed increase in energy for either metal clearly indicated strain-induced destabilization of the bulk W–Cu system under mutual lattice strain. Although these results were consistent with the known high $\Delta H_{mix}$ of the W–Cu system, they contradict our experimental observations of FCC-structured W–Cu nanoalloys, particularly in nanocrystals with high W contents ($W_{0.85}Cu_{0.15}$), which exhibited a stable FCC lattice

(**Fig. 1g**). This discrepancy suggested that surface effects likely played a paramount role in stabilizing the alloy structure at the nanoscale level.

The FCC (111) and BCC (110) surfaces represent the lowest-energy facets of the FCC and BCC lattices, respectively, in both the W and Cu crystals. The energies of these surfaces were compared as functions of both the W content and the average bond length. As shown in **Fig. 3c**, the surface energies of the BCC (110) (solid dots, lower panel) and FCC (111) (grid squares, upper panel) structures were plotted across the full W content (bond length) range for W (gray symbols) and Cu (orange symbols). Under mutual lattice strain, the surface energies of Cu decreased for both the BCC (110) and FCC (111) facets with increasing stretching of the Cu–Cu bonds. In contrast, the surface energies of W showed distinct trends. The FCC (111) surface energy rapidly decreased under mutual strain (compressed bond length) conditions, whereas the BCC (110) surface energy remained nearly constant throughout the wide content range. Thus, the alloying process altered both W and Cu into a common, intermediate alloy lattice, thereby inducing mutual strain.

Based on the DFT calculations, we proposed a mutual strain-induced surface energy lowering (MSI–SEL) mechanism. The mutual strain simultaneously stabilized the FCC (111) surfaces of both Cu and W, which partially offset the strain-induced destabilization of their bulk atoms and consequently promoted the formation of miscible W–Cu nanocrystals. This simultaneous surface stabilization process during size reduction was considered central to our proposed alloying/miscibility mechanism, in which MSI–SEL drove phase selection. Building upon the foundational energetic insights from bulk and surface analyses, we developed a quantitative energetic framework (detailed in the **Supplementary Information**) to predict the miscibility of W–Cu clusters as a function of both the crystal size and W–Cu composition ratio. The framework can be used to assess miscibility by comparing the calculated energies of the miscible solid–solution states and those of phase-separated heterodimer configurations as quantified by the energy difference ($E_m$–$E_{im}$) shown in **Fig. 3d**. For each composition, the shaded regions represent the range of energies derived by considering the plausible upper and lower bounds for the energy of the miscible phase, reflecting the inherent uncertainties in the model. These prediction results strongly suggest an MSI–SEL mechanism, indicating that the formation of FCC-structured W–Cu nanoalloys could be attributed primarily to the preferential stabilization of W in a metastable FCC structure.

**Understanding the atomically precise evolution of the miscibility and asymmetry shown in the W–Cu nanophase diagram**

In **Figures 4a** and **4b**, we present asymmetric and symmetric nanophase diagrams with the composition on the y-axis and the size on the x-axis (where the blue region indicates the miscible region and the orange region indicates the immiscible region). We selected five sets of data from the five compositions of W–Cu clusters and compiled them into a nanophase diagram (**Figure 4c**). This figure clearly reveals the asymmetric transition from immiscible (orange) to miscible (blue) as the number of atoms decreased and the W content increased: When the W content was 85%, miscibility was maintained in clusters with more than 6000 atoms, but when the W content reached

20%, miscibility was lost in clusters with approximately 650 atoms. This was the first experimental observation of an asymmetric nanophase. Although an asymmetric Au–Rh phase diagram under vacuum was predicted in earlier work, the adsorption of organic ligands during chemical synthesis preserved a symmetric profile in experiments.[10] We successfully obtained nanoalloys in small clusters of $W_{0.5}Cu_{0.5}$, $W_{0.65}Cu_{0.35}$, and $W_{0.85}Cu_{0.15}$. HAADF–STEM images of phase-separated clusters revealed significant separation of bright and dark atoms, which indicated the immiscibility of these clusters (**Supplementary Figure 8**).

When the lower limits in **Fig. 3d** and **Fig. 4c** were transformed into those shown in **Figure 4d**, the calculated nanophase diagrams that were derived from these energy differences suitably agreed with the experimental observations. Because the surface and bulk energies of Cu varied much more smoothly with strain than those of W did (**Fig. 3b** and **Fig. 3c**), the structural preference and miscibility were governed primarily by the competition between the FCC surface energy and the BCC bulk energy of W. The balance of this competition determined both the formation of FCC-structured W–Cu nanoalloys and their miscibility under mutual strain conditions. A simplified analysis suggested that a reduced nanocrystal size increased the importance of surface energy, thereby promoting the formation of miscible W–Cu nanocrystals. With respect to the composition, as the bulk energy of W increased sharply at lower W contents, a higher W content promoted alloying and enhanced the miscibility of the resulting nanocrystals, which is the reason for the formation of asymmetric nanophase diagrams. In addition, the formation of miscible nanoalloys could be attributable to a nonequilibrium generation process that involved rapid cooling during cluster formation, which, to some extent, regulated $T\Delta S_{mix}$ (**Fig. 1a**). The experimental and MSI-SEL mechanism results highlight the critical roles of both the number of atoms and the composition in governing the asymmetric miscibility behaviors of W–Cu nanoalloys and provide a quantitative framework for understanding the nanophase behaviors of immiscible elements.

**Conclusions**

We systematically investigated the alloying of an oxidation-prone W–Cu system with high mixing enthalpy via an atomically size-selected cluster beam source. We demonstrated the presence of an FCC structure in $W_{0.85}Cu_{0.15}$ clusters that contained approximately 2280 atoms, with a lattice constant of 3.88 Å. For the first time, an unexpected asymmetric nanophase diagram was observed experimentally, in which W-rich compositions favor mixing, and the critical size is approximately 6000 atoms. We attributed this trend to an MSI–SEL mechanism. The DFT calculation results revealed that reducing the cluster size increased the relative importance of surface energy, whereas the mutual strain simultaneously decreased the FCC (111) surface energies of both W and Cu, thereby facilitating the formation of miscible W–Cu nanocrystals. Collectively, our work offers a preparation strategy that can guide the generation of binary and even high-entropy alloy systems with atomical precision. At the mechanistic level, this study presents a new MSI-SEL mechanism for understanding the asymmetric miscibility of binary systems as a function of the number of atoms and composition, thus providing predictive insights into the control of miscibility in the design of future new high-performance nonequilibrium phase alloys.

## Methods

### Materials

The targets used for magnetron sputtering in the cluster source were purchased from Hebei Qinbang New Material Technology Co., Ltd., with 99.99% purity. Au TEM grids coated with carbon film were purchased from Beijing Zhongjingkeyi Technology Co., Ltd. Single-sided polished N-doped Si substrates that were sliced into 10 × 10 mm$^2$ pieces were purchased from Shenzhen Shunsheng Electronic Science and Technology Company. A FISCHIONE Ti tip vacuum transfer tomography holder was purchased from Shanghai Winner International Trading Co., Ltd.

### Preparation of atomically precise W–Cu nanoalloys and clusters

The mass-selected clusters were generated via a magnetron sputtering gas-phase condensation cluster beam source. W–Cu clusters with various compositions were obtained by adjusting the composition of the target material and regulating the physical parameters of the cluster source.[48] A time-of-flight mass selector was used to select clusters with specific mass numbers and offered a mass resolution of M/ΔM≈10. The mass-selected metal clusters were focused into the deposition chamber under high vacuum conditions (10$^{-5}$ Pa) and were deposited onto N-doped silicon wafers with adjustable deposition energy to ensure soft landing conditions. Au TEM grids coated with carbon films were employed to simultaneously collect the raw clusters.

### XPS data acquisition

XPS spectra of the $W_{0.85}Cu_{0.15}$ cluster were obtained via X-ray photoelectron spectroscopy (ESCALAB 250XI, Thermo Fisher). In-depth analyses of the valence states of the samples were performed via an etching process with the following parameters: the energy of the argon ion gun was 2000 V with a working current of 0.3 μA, the etching area was 600×600 μm$^2$, and the etching rate was 1.0 nm/s on the basis of $Ta_2O_5$. Each etching process lasted 5 s.

**STEM, 4D-STEM and electron diffraction characterization**

EDS and STEM characterization analyses of the clusters were performed with an FEI Titan Themis Z TEM instrument and an FEI Titan 60–300 TEM instrument. The FEI Titan Themis Z TEM instrument was operated at an acceleration voltage of 200 kV and was equipped with a probe spherical aberration corrector and a SuperXG23 detector. The FEI Titan 60–300 TEM instrument was operated at an acceleration voltage of 300 kV and was equipped with double aberration correctors and a SuperXG1 detector. The L peaks of W and the K peaks of Cu in the EDS spectra were used for elemental mapping and compositional quantification of the clusters. Each EDS map was built on the basis of the accumulation of 150 frames with a pixel dwell time of 8 μs.

HAADF–STEM images were acquired with a high-angle annular dark-field detector with collection angles that ranged from 44 to 200 mrad for the FEI Titan Themis Z instrument and from 37 to 200 mrad for the FEI Titan 60–300 instrument. The beam convergence was 25.6 mrad for the FEI Titan Themis Z instrument and 22.5 mrad for the FEI Titan 60–300 instrument. The beam current was set to 40 pA for the imaging of clusters. All the HAADF–STEM images were recorded in a 1,024×1,024 pixel$^2$ format with a pixel dwell time of 8 μs.

Nanobeam electron diffraction (NBED) patterns were obtained with an FEI Titan Themis Z transmission electron microscope that was operated at 300 kV with a probe spherical aberration corrector. The electron beam was directed perpendicular to the plane of the carbon film in nanoprobe mode. A low convergence angle of 0.3 mrad and a camera length of 910 mm were employed to ensure complete separation of the diffraction points from the nanoparticles. The 4D-STEM data were acquired using an electron microscope pixel array detector (EMPAD) at a frame rate of 1000 frames/sec, with each scan position captured in approximately 2-nm steps to construct the 4D dataset. Nanoparticles with low-index zone axes aligned along the electron beam direction were selected using the open-source py4DSTEM script to extract the NBED images.

Selected area electron diffraction (SAED) images were captured using an FEI Titan Themis Z TEM instrument that was operated at 200 kV. For SAED acquisition, we used a selected-area aperture with a physical diameter of 40 μm.

**Random walk model-based STEM image analysis**

Given the extremely low signal intensity from clusters that contained only a few hundred atoms during EDS acquisition (**Supplementary Figure 9**) and the structural changes that were induced by electron beam irradiation (**Supplementary Figure 10**), we developed an algorithm that is based on the random walk model for detecting bright regions in STEM images to identify W-enriched areas. The algorithm, which was implemented in MATLAB v.R2024b, simulates random walk processes and integrates image processing techniques to identify clusters of high-intensity pixels.

**DFT calculations**

Calculations were performed by using the generalized gradient approximation (GGA) in the Perdew–Burke–Ernzerhof revised functional for solids (PBEsol).[49] To evaluate the exchange–correlation potential, the projector augmented wave (PAW) method[50] and a plane wave basis set were implemented in the Vienna ab initio simulation package (VASP).[51]

Bulk $W_xCu_{1-x}$ alloys with various compositions (x = 0.85, 0.8, 0.7, 0.65, 0.35, 0.3, or 0.2) were modeled via the virtual crystal approximation (VCA) method, where W and Cu pseudopotentials were linearly combined according to the desired composition.[52] Both BCC and FCC crystal structures were considered. An energy cutoff of 700 eV was used for the plane wave basis set in the VCA calculations. A k-mesh of dimensions 15×15×15 was used to sample the first Brillouin zone of the bulk $W_xCu_{1-x}$ alloys. The atomic positions and cell parameters were fully relaxed until the residual force on each atom was less than 0.01 eV Å$^{-1}$.

Surface energies were computed with symmetric slab models that contained 9 atomic layers, with a vacuum region of at least 15 Å along the direction perpendicular to the surface to prevent interslab interactions. The surface energy ($\gamma$) was defined as $\gamma = \frac{E_{slab} - NE_{bulk}}{2A}$, where $E_{slab}$ represents the total energy of the relaxed slab, $N$ represents the number of atoms in the slab, $E_{bulk}$ represents the energy per atom of the corresponding bulk material, and $A$ represents the surface area of one side of the slab. An energy cutoff of 700 eV was used for the plane wave basis set in the surface energy calculations. A k-mesh of dimensions 12×12×1 was used to sample the first Brillouin zone of each slab model. During structural relaxation, the atoms in the three central layers were fixed to their bulk positions, whereas the atoms in the uppermost three layers and the lowermost three layers were allowed to relax until the residual force on each atom was less than 0.01 eV Å$^{-1}$.

For the nanocrystal energy calculations, each truncated cubic nanocrystal was placed in a cubic supercell with dimensions of 40×40×40 Å$^3$, which ensured a minimum vacuum separation of 15 Å between the nanocrystal and its periodic images to minimize spurious interactions. A plane wave energy cutoff of 350 eV was used, and the Brillouin zone was sampled only at the Γ point (1×1×1 k-mesh). During structural optimization, all the surface atoms of the nanocrystal were allowed to relax until the residual forces on each atom were less than 0.02 eV Å$^{-1}$, while the atoms in the bulk were fixed to their bulk positions.

**Acknowledgments**

The authors are grateful to the National Graphene Products Quality Inspection and Testing Center in Wuxi, China.

**Funding**

This work was supported financially by the National Key R&D Program of China (Nos. 2022YFA1402404, 2024YFA1409600, and 2023YFA1406500), the National Natural Science Foundation of China (Nos. 92161201, T2221003, 12422410, 12474272, 12374043, 12274208, 12025404, 52025012, 12274202, 12374253, 92477205, 12274202 and 52461160327), the Natural Science Foundation of Jiangsu Province (No. BK20243013, No. BK20240166, No. BK20230079, and No. BK20233001), the Key R&D Program of Liaoning Province (2023021341-JH2/1018), and the Fundamental Research Funds for the Central Universities and the Research Funds of Renmin University of China (Grant No. 22XNKJ30). Calculations were performed at the Physics Lab of High-Performance Computing (PLHPC) and the Public Computing Cloud (PCC) of Renmin University of China.

**Author contributions**

F. S. conceived and supervised the entire study. Y. Z. was in charge of sample preparation. Y. Z., Y. L., D. S., Z. P. and Y. R. carried out the sample measurements. Y. Z., Z. Z., S. L., Y. L., D. S., C. W., K. H., D. W., Y. D. and B. G. participated in the experimental data analysis. W. Z., L. Z., J. G., W. G., J. Z. and W. J. were in charge of the DFT calculations. F. S., Y. Z., W. Z. and W. J. wrote the manuscript. All the authors discussed the results and commented on the manuscript.

Y. Z. and W. Z. contributed equally to this work.

**Competing interests**

The authors declare that they have no competing interests.

**Figure captions**

**Figure 1. Successful preparation and structural characterization of miscible $W_{0.85}Cu_{0.15}$ nanoalloys.** **(a)** Schematic illustration of the gas phase condensation process for mass-selected cluster generation, which includes two parts: cluster generation and rapid cooling, and precise cluster mass selection. **(b)** Bulk phase diagram of the W–Cu binary system, where Cu (orange) adopts an FCC structure and W (gray) adopts a BCC structure. **(c)** EDS spectrum of $W_{0.85}Cu_{0.15}$ clusters, which shows the presence of W and Cu peaks. **(d)–(g)** Characterization of $W_{0.85}Cu_{0.15}$ clusters with a mass of 378488 a.m.u.: **(d)** low-magnification STEM-HAADF image, **(e)** SAED pattern indexed by using the FCC structure parameters, **(f)** high-magnification STEM–HAADF image, and **(g)** NBED pattern, which shows an FCC structure with a [110] zone axis.

**Figure 2. Investigation of W–Cu cluster evolution as a function of the W content and number of atoms as well as their miscibility (the scale bar represents 2 nm).** **(a)** EDS spectra of $W_xCu_{1-x}$ (x= 0.65, 0.5, 0.3, or 0.2) clusters, with compositions determined by EDS measurements. **(b)–(c)** EDS mapping and line scans of $W_{0.85}Cu_{0.15}$ and $W_{0.2}Cu_{0.8}$ clusters with masses of 935272 a.m.u. **(d)–(e)** Random walk-based detection of bright regions in STEM images (top: HAADF–STEM images; bottom: W-enriched regions): **(d)** miscibility as a function of the number of atoms for $W_{0.5}Cu_{0.5}$ clusters of five distinct sizes (from left to right: mass numbers of 56856, 103224, 169832, 378488, and 935272 a.m.u.) and **(e)** miscibility as a function of composition

for clusters with a mass of 169832 a.m.u. (from left to right: $W_xCu_{1-x}$, x=0.85, 0.65, 0.5, 0.3, and 0.2).

**Figure 3. DFT calculations of the energetic factors that govern the miscibility and phase stability of W–Cu clusters.** **(a)** Site density (fraction of atoms at the bulk, surface, edge, or corner sites) as a function of the total number of atoms in a truncated cubic nanoalloy model. The inset shows a schematic illustration of a truncated cube nanocrystal, with the (100) and (111) facets indicated. **(b)** Calculated bulk energy per atom ($E_{bulk}$) for pure W (gray) and pure Cu (orange) in both FCC (grid squares) and BCC (solid dots) lattices. The energies are plotted against the W content, and the corresponding alloy bond length is derived from the VCA shown on the top axis. **(c)** Calculated surface energies for the lowest-energy FCC (111) (upper panel, grid squares) and BCC (110) (lower panel, solid dots) facets for both W (gray) and Cu (orange). The surface energies are plotted as a function of W content and the VCA-derived bond length. **(d)** Calculated energy difference per atom ($E_m$ - $E_{im}$) between the miscible ($E_m$) and immiscible ($E_{im}$, heterodimer) states as a function of the total number of atoms per nanoparticle for W–Cu clusters of various compositions. The shaded regions for each composition ($W_xCu_{1-x}$, x=0.85, 0.65, 0.5, 0.3, 0.2) represent the energy range derived from considering two limiting models for $E_m$: an energetically optimized solid solution and a linear combination reference that is based on pure W and Cu nanocrystals (see Methods for details). Negative values indicate a preference for the miscible state.

**Figure 4. Asymmetry in the experimental and computational nanophase diagrams of W–Cu clusters. (a) (b)** Schematic illustrations of asymmetric and symmetric miscibility phases with the composition as the y-axis and the size as the x-axis shown separately (the blue area represents miscibility, whereas the orange area represents immiscibility). **(c)** W–Cu clusters with five masses of data from the five compositions that were statistically analyzed to summarize the nanophase diagram of W–Cu clusters as a function of size (mass and number of atoms) and composition (the blue region represents miscibility, whereas the orange region represents immiscibility). **(d)** Combination and comparison of the lower limit of the evolution of miscibility shown in **Fig. 3d** with the experimental points in **Fig. 4a**. (DFT: Orange indicates immiscible, and blue indicates miscible; experiment: ★ indicates immiscibility, and ● indicates miscibility).

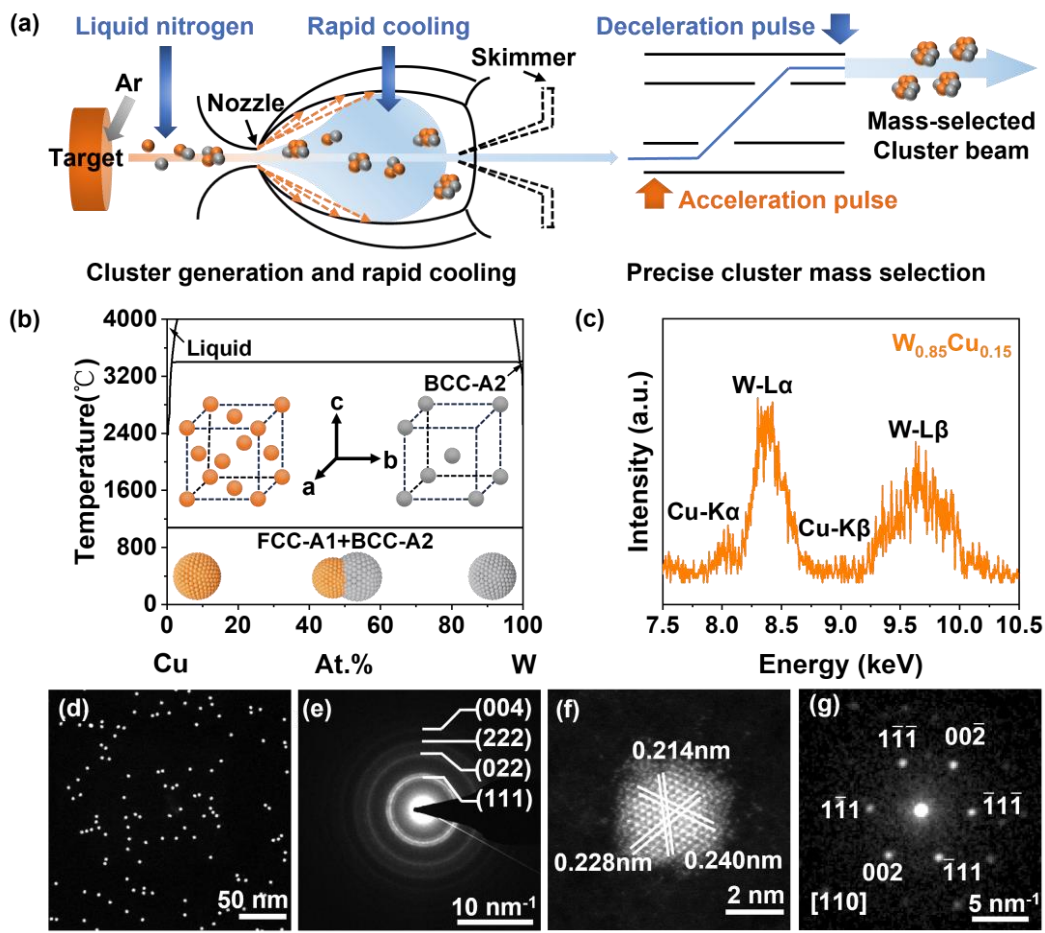

**Figure 1. Successful preparation and structural characterization of miscible W$_{0.85}$Cu$_{0.15}$ nanoalloys.**

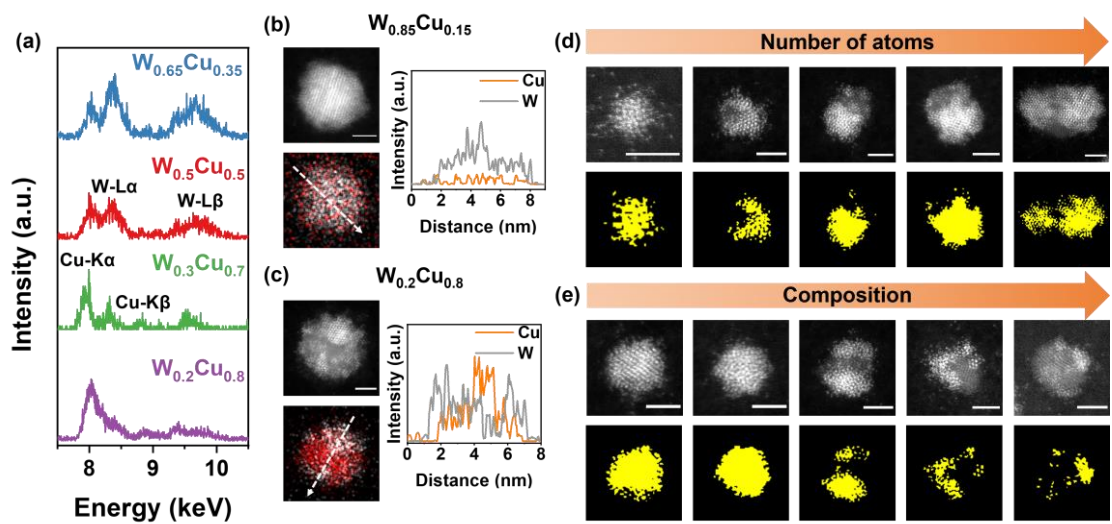

**Figure 2.** Investigation of W–Cu cluster evolution as a function of the W content and number of atoms as well as their miscibility (the scale bar represents 2 nm).

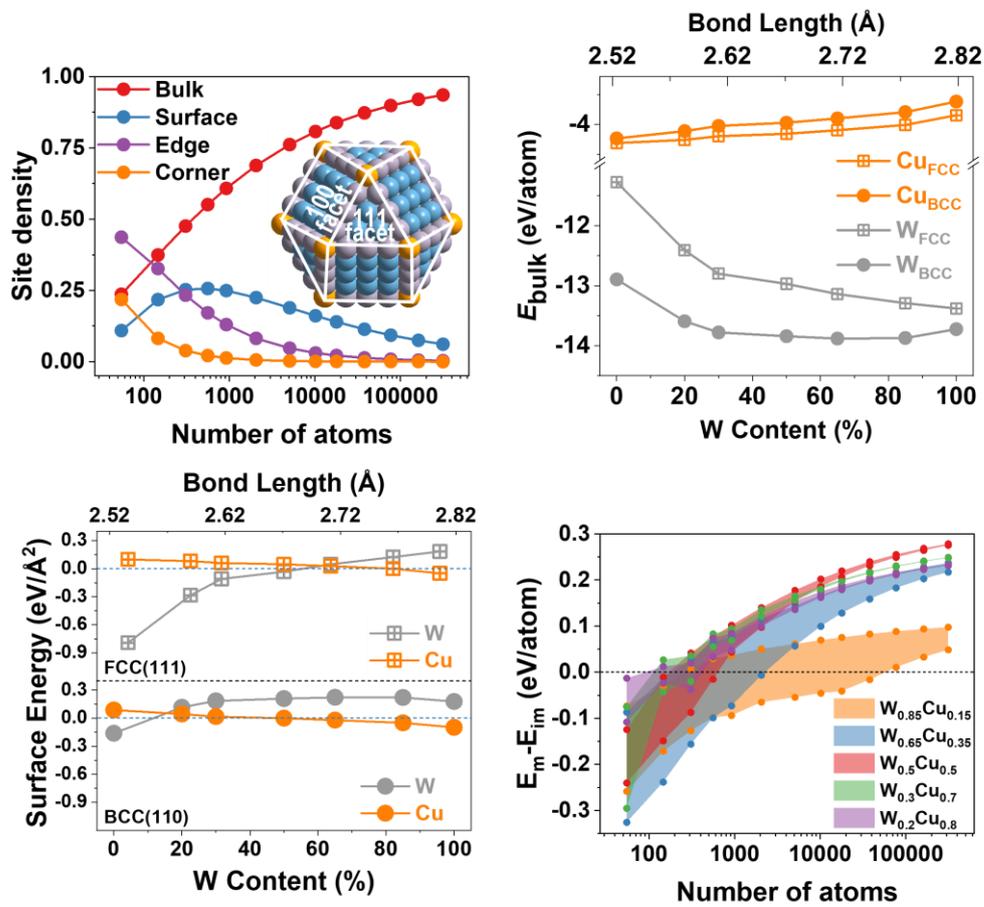

**Figure 3.** DFT calculation results for the energetic factors that govern the miscibility and phase stability of W–Cu clusters.

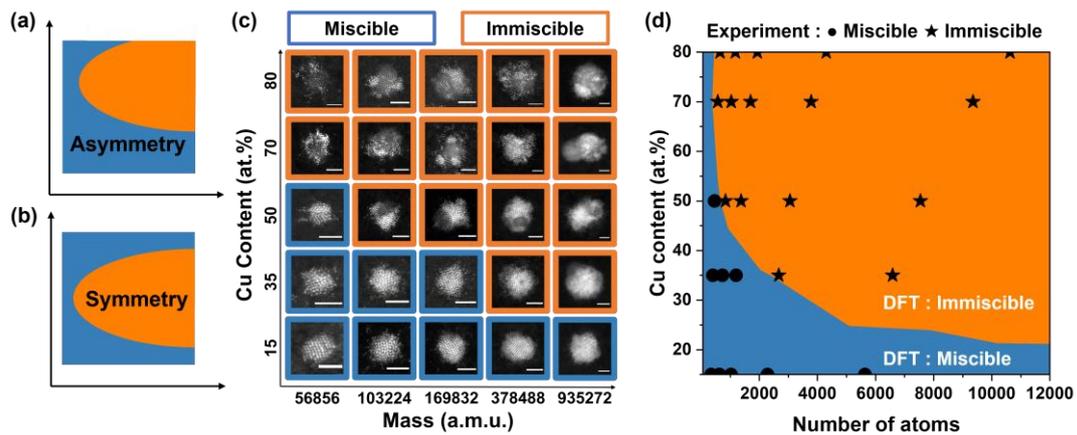

**Figure 4.** Asymmetry in the experimental and computational nanophase diagrams of W–Cu clusters.